\documentstyle[12pt]{article}
\newcommand{\beq}{\begin{equation}}
\newcommand{\eeq}{\end{equation}}
\newcommand{\beqa}{\begin{eqnarray}}
\newcommand{\eeqa}{\end{eqnarray}}
\newcommand{\ba}{\begin{array}}
\newcommand{\ea}{\end{array}}
\newcommand{\CR}{\nonumber \\}
\newcommand{\pa}{\partial}
\newcommand{\A}{\alpha}
\newcommand{\B}{\beta}
\newcommand{\D}{\delta}          

\newcommand{\E}{\epsilon}
\newcommand{\La}{\Lambda}
\newcommand{\p}{\Phi}
\newcommand{\bra}{\langle}
\newcommand{\ket}{\rangle}
\newcommand{\lm}{\lambda}
\newcommand{\s}{\sigma}
\newcommand{\tQ}{\tilde{Q}}

\begin{document}

\begin{titlepage}
\null
\begin{flushright} 
9706076  \\
UTHEP-316 \\
June, 1997
\end{flushright}
\vspace{0.5cm} 
\begin{center}
{\Large \bf
ADE Confining Phase Superpotentials
\par}
\lineskip .75em
\vskip2.5cm
\normalsize
{\large Seiji Terashima and Sung-Kil Yang} 
\vskip 1.5em
{\it Institute of Physics, University of Tsukuba, Ibaraki 305, Japan}
\vskip3cm
{\bf Abstract}
\end{center} \par
We obtain a low-energy effective superpotential for a phase with a 
single confined photon in $N=1$ gauge theory with an adjoint matter with ADE
gauge groups. The expectation values of gauge invariants built out of the 
adjoint field parametrize the singularities of moduli space of the Coulomb 
phase. The result can be used to derive the $N=2$ curve in the form of 
a foliation over ${\bf CP}^1$. Our $N=1$ theory exhibits non-trivial fixed 
points which naturally inherit the properties of the ADE classification of 
$N=2$ superconformal field theories in four dimensions. We also discuss how to
include matter hypermultiplets toward deriving the Riemann surface
which describes $N=2$ QCD with exceptional gauge groups.
\end{titlepage}

\baselineskip=0.7cm

\section{Introduction}

Recently $N=2$ supersymmetry has played a profound role in understanding 
strong-coupling dynamics of gauge and string theories in various dimensions
\cite{review}.
In four dimensions the vacuum structure of the $N=2$ Coulomb phase is 
described in terms of the Riemann surface \cite{SeWi}. 
This geometry of $N=2$ Yang-Mills 
theory is called Seiberg-Witten (SW) geometry. 

An interesting approach to the issue of SW geometry is based on $N=1$
supersymmetry.
When $N=2$ theory is perturbed by a tree-level superpotential
explicitly breaking $N=2$ to $N=1$ supersymmetry it is observed that only
the singularities of moduli space where monopoles or dyons become massless
remain as the $N=1$ vacua \cite{SeWi}. 
Thus, studying the low-energy properties of 
$N=1$ Yang-Mills theory with an adjoint matter field with a tree-level
superpotential chosen properly one may derive the singular loci of
$N=2$ moduli space \cite{InSe},\cite{ElFoGiRa},\cite{ElFoGiRa2}. 
For this purpose, using the ``integrating-in'' technique \cite{In},\cite{InSe},
Elitzur et al. have developed a method
of $N=1$ confining phase superpotential by focusing on a phase with a single
confined photon \cite{ElFoGiInRa}. This approach has now been extended 
for supersymmetric Yang-Mills
theory as well as QCD with all classical gauge groups
\cite{TeYa},\cite{Ki},\cite{KiTeYa},\cite{GPR}. 
In the case of
exceptional gauge groups only $G_2$ gauge theory has been analyzed
so far \cite{LaPiGi}.

In this article we wish to show that the method described above applies in a
unified way in determining the singularity structure of moduli space of the
Coulomb phase in
supersymmetric gauge theories with ADE gauge groups. Not only the classical
case of $A_r, D_r$ groups but the exceptional case of $E_6, E_7, E_8$ 
groups can be treated on an equal footing since our discussion is based on
the fundamental properties of the root system of the simply-laced Lie
algebras.

For exceptional gauge groups there has accumulated considerable evidence that
SW geometry is not realized by hyperelliptic curves 
\cite{MaWa},\cite{LaPiGi},\cite{LW},\cite{WY},\cite{Ito}. 
In fact we will see that
the Riemann surface described as a foliation over ${\bf CP}^1$ satisfies the
singularity conditions we obtain from the $N=1$ confining phase superpotential.
This Riemann surface is not of hyperelliptic type for exceptional gauge groups.

In sect.2 we derive the low-energy effective superpotential which is used to
determine the singularities of moduli space in $N=1$ theory, which in turn
enables us to construct $N=2$ SW geometry for ADE gauge groups. 
In sect.3 the ADE series of $N=1$ superconformal field theories realized at 
particular vacuum in the Coulomb phase is discussed. In sect.4 we study how
the results in sect.2 obtained for $N=1$ pure Yang-Mills theory with an adjoint
matter is extended so as to include chiral matter multiplets. Finally in 
sect.5 we draw our conclusions.

\section{N=1 theory with an adjoint matter}

Let us consider $N=1$ supersymmetric gauge theory
with an adjoint matter $\Phi$.
We assume that the gauge group $G$ is simple and simply-laced, namely, 
$G$ is of ADE type. Our purpose in this section is to show that, under 
appropriate ansatz, the low-energy effective superpotential for 
the Coulomb phase is obtained in a unified way for all ADE gauge groups just 
by using the fundamental properties of the root system $\Delta$. Our notation
for the root system is as follows. The simple roots of $G$ are denoted as
$\alpha_i$ where $1 \leq i \leq r$ with $r$ being the rank of $G$. Any root
is decomposed as $\alpha =\sum_{i=1}^{r} a^i \alpha_i$. The component indices
are lowered by $a_i=\sum_{j=1}^{r} A_{ij} a^j$ where $A_{ij}$ is the ADE
Cartan matrix. The inner product of two roots $\alpha$, $\beta$ are then
defined by
\beq
\A \cdot \B = \sum_{i=1}^{r} a^i b_i = \sum_{i,j=1}^{r} a^i A_{ij} b^j,
\eeq
where $\B=\sum_{i=1}^{r} b^i \A_i$. For ADE all roots have the equal norm and
we normalize $\alpha^2=2$.

In our $N=1$ theory we take a tree-level superpotential 
\beq
W=\sum_{k=1}^{r} g_{k} u_{k}(\Phi),
\label{tree}
\eeq
where $u_k$ is the $k$-th Casimir of $G$ constructed from $\Phi$
and $g_k$ are coupling constants. The mass dimension of $u_k$ is $e_k+1$
with $e_k$ being the $k$-th exponent of $G$.
When $g_k=0$ $\Phi$ is considered as the
chiral field in the $N=2$ vector multiplet and we have $N=2$ ADE supersymmetric
gauge theory. 

We first make a classical analysis of the theory with the superpotential 
(\ref{tree}).
The classical vacua are determined by the equation of motion
$\frac{\pa W}{\pa \Phi}=0$ and the $D$-term equation.
Due to the $D$-term equation,
we can restrict $\Phi$ to take the values in the Cartan subalgebra
by the gauge rotation.
We denote the vector in the Cartan subalgebra corresponding to the
classical value of $\Phi$ as $a=\sum_{i=1}^{r} a^i \A_i$.
Then the superpotential becomes
\beq	
W(a)=\sum_{k=1}^{r} g_{k} u_{k}(a),
\label{tree2}
\eeq
and the equation of motion reads
\beq
\frac{\pa W(a)}{\pa a^i}=
\sum_{k=1}^{r} g_{k} \frac{\pa u_{k}(a)}{\pa a^i}=0.
\label{eq1}
\eeq
For $g_k \not\equiv 0$ we must have
\beq
J(a) \equiv {\rm det} \left(  \frac{\pa u_{j}(a)}{\pa a^i} \right) =0.
\label{Jzero}
\eeq
According to \cite{Hu} it follows that
\beq
J(a) =c_1 \prod_{\A \in \Delta^+} a \cdot \A ,
\label{J}
\eeq
where $\Delta^+$ is a set of positive roots and $c_1$ is a certain constant. 

The condition $J(a)=0$ means that the vector $a$ hits a wall of the
Weyl chamber and there occurs enhanced gauge symmetry. Suppose that the vector
$a$ is orthogonal to a root, say, $\A_1$
\beq
a \cdot \A_1=0,
\label{su2sol}
\eeq
where $\A_1$ may be taken to be a simple root.
In this case we have the unbroken gauge group
$SU(2) \times U(1)^{r-1}$ where the $SU(2)$ factor is spanned by 
$\{ \A_1 \cdot H, E_{\A_1}, E_{-\A_1} \}$ in the Cartan-Weyl basis.
If some other factors of $J$ vanish besides $a \cdot \A_1$
the gauge group is further enhanced from $SU(2)$. 
Since $SU(2) \times U(1)^{r-1}$
is the most generic unbroken gauge group we shall restrict ourselves to
this case in what follows.

We remark here that there is the case in which the $SU(2) \times U(1)^{r-1}$
vacuum is not generic. As a simple, but instructive example consider
$SU(4)$ theory. Casimirs are taken to be
\beqa
u_1 &=& {1\over 2} {\rm Tr}\, \Phi^2, \CR
u_2 &=& {1\over 3} {\rm Tr}\, \Phi^3, \CR
u_3 &=& {1\over 4} {\rm Tr}\, \Phi^4 
-\alpha \left( {1\over 2}{\rm Tr}\, \Phi^2 \right)^2, 
\eeqa
where $\alpha$ is an arbitrary constant. If we set $\alpha =1/2$ it is observed
that the $SU(2) \times U(1)^2$ vacuum exists only for the special values
of coupling constants, $(g_2/g_3)^2=g_1/g_3$. Thus, for $\alpha =1/2$, the
$SU(2) \times U(1)^2$ vacuum is not generic though it does so for 
$\alpha \not= 1/2$. This points out that we have to choose the appropriate
basis for Casimirs when writing down (\ref{tree}) to have the 
$SU(2) \times U(1)^{r-1}$ vacuum generically \cite{TeYa}.

Now we assume that
there is no mixing between the $SU(2) \times U(1)^{r-1}$ vacuum and
other vacua with different unbroken gauge groups.
According to the arguments of \cite{InSe2},
we should not consider the broken gauge group instantons.
We thus expect that there is only perturbative effect 
in the energy scale above the scale $\La_{YM}$ of the 
low-energy effective $N=1$ supersymmetric 
$SU(2)$ Yang-Mills theory.

\subsection{Higgs mass}

Our next task is to evaluate the Higgs scale
associated with the spontaneous breaking of the 
gauge group $G$ to $SU(2) \times U(1)^{r-1}$.
For this purpose we decompose the adjoint representation of $G$ to 
irreducible representations of $SU(2)$. We fix the $SU(2)$ direction by taking
a simple root $\alpha_1$. It is clear that the spin $j$ of every representation
obtained in this decomposition satisfies $j \leq 1$ since all roots have the
same norm and the $SU(2)$ raising (or lowering) operator shifts a root $\alpha$
to $\alpha +\alpha_1$ (or $\alpha -\alpha_1$). The fact that there is no
degeneration of roots indicates that the $j=1$ multiplet has the
roots $(\alpha_1,0,-\alpha_1)$ corresponding to the unbroken $SU(2)$
generators. The roots orthogonal to $\alpha_1$ represent the $j=0$ multiplets.
The $j=1/2$ multiplets have the roots $\alpha$ obeying 
$\A \cdot \A_1=\pm 1$. Let us define a set of these roots by
$\Delta_d=\{ \alpha | \alpha \in \Delta,\, \alpha \cdot \alpha_1 =\pm 1 \}$.
For each root $\alpha \in \Delta_d$ there appears a massive gauge boson.
These massive bosons pair up in $SU(2)$ doublets with weights $(\alpha,
\alpha \pm \alpha_1)$ which indeed have the same mass $|a\cdot \alpha |=
|a\cdot (\alpha \pm \alpha_1)|$ since $a\cdot \alpha_1=0$.

We now integrate out the fields that become massive by the Higgs mechanism.
The massless $U(1)^{r-1}$ degrees of freedom are decoupled. 
The resulting theory characterized by the scale $\La_H$ is $N=1$ 
$SU(2)$ theory with an adjoint chiral multiplet.
The Higgs scale $\La_H$ is related to the high-energy scale $\La$ through
the scale matching relation
\beq
\La^{2 h}= \La_H^{2\cdot 2} \left( \prod_{\B \in \Delta_d,\, \B >0} 
a \cdot \B \right)^\ell,
\label{summaa} 
\eeq
where $2h=4+\ell n_d/2$, $n_d$ is the number of elements in $\Delta_d$ 
and $h$ stands for the 
dual Coxeter number of $G$; $h=r+1, 2r-2, 12, 18,30$ for $G=A_r, D_r, 
E_6, E_7, E_8$ respectively. The reason for $\beta >0$
in (\ref{summaa}) is that weights $(\beta ,\beta \pm \alpha_1)$ of an
$SU(2)$ doublet are either both positive or both negative 
since $\alpha_1$ is the simple
root, and gauge bosons associated with $\beta <0$ and $\beta >0$ have the
same contribution to the relation (\ref{summaa}).

To fix $\ell$ we calculate $n_d$ by evaluating the quadratic Casimir $C_2$
of the adjoint representation in the following way. Taking hermitian generators
we express $C_2$ in terms of the structure constants $f_{abc}$ through
$\sum_{a,b}f_{abc}f_{abc'}=-C_2 \, \delta_{cc'}$. From the commutation 
relation 
$[\alpha_1\cdot H, E_\alpha ]=(\alpha_1 \cdot \alpha) E_\alpha$ one can check
\beq
C_2={1\over 2}\sum_{\A \in \Delta} ( \A_1 \cdot \A )^2
={1\over 2} \left( \sum_{\A \in \Delta_d} ( \A_1 \cdot \A )^2 + 
2 (\alpha_1 \cdot \alpha_1)^2 \right)= {1\over 2} \left ( n_d +8 \right).
\eeq
On the other hand, the dual Coxeter number $h$ is given by
$h=C_2/\theta^2$ with $\theta$ being the highest root. We thus find
\beq
n_d = 4 (h-2)
\label{numbd}
\eeq
and (\ref{summaa}) becomes
\beq
\La^{2 h}= \La_H^{2 \cdot 2} \prod_{\B \in \Delta_d,\, \B >0} a \cdot \B .
\label{summa}
\eeq

\subsection{Adjoint mass}

After integrating out the massive fields due to the Higgs mechanism we are
left with $N=1$ $SU(2)$ theory with the massive adjoint.
In order to evaluate the mass of the adjoint chiral multiplet $\Phi$
we need to clarify some properties of Casimirs.
Let $\s_\B$ be an element of the Weyl group of $G$ 
specified by a root $\B= \sum_{i=1}^r b^i \A_i$.
The Weyl transformation of a root $\alpha$ is given by
\beq
\s_\B (\A) = \A - (\A \cdot \B) \B .
\label{Weyl}
\eeq
When $\s_\B$ acts on the Higgs v.e.v. vector $a=\sum_{i=1}^r a^i \A_i$ we have
\beq
{a'}^i = \sum_{j=1}^r {S_\B}^i_{\; j} \; a^j, \hskip5mm
{S_\B}^i_{\; j} \equiv \D^i_{\; j}-b^i b_j,
\eeq
where $\s_\B (a)=\sum_{i=1}^r a'^i \A_i$. 
Since the Casimirs $u_k(a)$ are Weyl invariants it is obvious to see
\beq
\frac{\pa}{\pa a^i} u_k(a) = \frac{\pa}{\pa a^i} u_k(a')=
\sum_{j=1}^r {S_\B}^j_{\; i} 
\left. \left( \frac{\pa}{\pa {a}^j} u_k(a) \right) \right|_{a \rightarrow a'}.
\eeq
Let $\bar a$ be a particular v.e.v. which is fixed under the action 
of $\s_\B$, then we find the identity
\beq
\left. \sum_{j=1}^r \left( \D^j_{\; i} - {S_\B}^j_{\; i} \right)
\frac{\pa}{\pa {a}^j} u_k(a) \right|_{a=\bar a} =0
\eeq
for all $i$, and thus
\beq
\left. \sum_{j=1}^r b_j \frac{\pa}{\pa {a}^j} u_k(a) \right|_{a=\bar a} =0.
\eeq
This implies that for any v.e.v. vector $a$ 
and root $\beta$ we can write down 
\beq
\sum_{j=1}^r b^j \frac{\pa}{\pa {a}^j} u_k(a)= (a \cdot \B) \; u_k^\B (a),
\label{diff}
\eeq
where $u_k^\B (a)$ is some polynomial of $a^i$.
If we set $\beta =\alpha_i$, a simple root, we obtain a useful formula
\beq
\frac{\pa}{\pa {a}^i} u_k(a)= a_i  \; u_k^{\A_i} (a).
\label{diff2}
\eeq

As an immediate application of the above results, for instance, we point out
that (\ref{J}) is derived from (\ref{diff}) and the fact that 
the mass dimension of $J(a)$ is given by
\beq
\sum_{k=1}^r e_k={1\over 2}\, ({\rm dim}\,  G-r),
\eeq
where $e_k$ is the $k$-th exponent of $G$.

Let us further discuss the properties of $u_k^{\A_j} (a)$. Define $D_{mn}$ as
\beq
D_{mn}  \equiv (-1)^{n+m} {\rm det} 
\left( \frac{\pa u_{\tilde{j}}(a)}{\pa a^{\tilde{i}}} \right),
\hskip5mm 1 \leq m,n \leq r ,
\eeq
where $1 \leq \tilde{i},\tilde{j} \leq r$ with $\tilde{i} \not= m,
\tilde{j} \not= n$,
then $D_{1n}$ is a homogeneous polynomial of $a^i$ with the mass dimension
$\sum_{k=1}^{r} e_k -e_n$.
We also denote $\Delta_e$ as a set of positive roots where $\A_1$ and 
$SU(2)$ doublet roots $\A$ with $\A+\A_1 \not\in \Delta^+$ are excluded.
If we set $a_1=0$ and $a \cdot \B=0$ where $\B$ is 
any root in $\Delta_e$
we see $D_{1n}=0$ from the identity (\ref{diff}).
Consequently we can expand 
\beq
D_{1n}=h_n(a) \prod_{\B \in \Delta_e} (a \cdot \B)+a_1 f_n(a),
\eeq
where $h_n(a)$, $f_n(a)$ are polynomials of $a_i$. In particular
\beq
D_{1r}=c_2 \prod_{\B \in \Delta_e} (a \cdot \B)+a_1 f_r(a),
\label{Dexpr}
\eeq
where $c_2$ is a constant. Notice that the first term on the rhs has the 
correct mass dimension since the number of roots in $\Delta_e$ reads
\beq
{1\over 2}\, ({\rm dim}\,  G-r) -1-{n_d\over 4}=\sum_{k=1}^r e_k -(h-1),
\eeq
where we have used (\ref{numbd}) and $e_r=h-1$.

We are now ready to evaluate the 
mass of $\Phi$ in intermediate $SU(2)$ theory.
The fluctuation of $W(a)$ around the classical vacuum yields the adjoint mass.
To find the mass relevant for the scale matching we should only 
consider the components of $\Phi$ which are coupled to the 
unbroken $SU(2)$. The mass $M_\Phi$ of these components is then given by
\beq
2 M_\Phi=\left. \frac{\pa^2}{(\pa a^1)^2} W(a) \right |
=\left. \frac{\pa}{\pa a^1} (a_1 W_1) \right | =
\left. \left( a_1 \frac{\pa}{\pa a^1} W_1+2 W_1 \right) \right| 
=\left. 2 W_1 \right| ,
\label{massofadj}
\eeq
where $W_1=(\sum_{k=1}^r g_k u_k^{\A_1}) (a) $
and $a^i$ are understood as solutions of the equation of motion (\ref{eq1}).

To proceed further it is convenient to rewrite the equation motion (\ref{eq1})
and the vacuum condition (\ref{su2sol}) with the simple root 
$\alpha_1$ as follows:
\beqa
g_1 : g_2 : \cdots : g_r &=& D_{11} : D_{12} : \cdots : D_{1r},
\CR
a_1& =& 0.
\label{eqmoratio}
\eeqa
The solutions of these equations are expressed as functions 
of the ratio $g_i/g_r$. Then we notice that $J(a)$ defined in 
(\ref{Jzero}) turns out to be
\beq
J=\sum_{k=1}^r \frac{\pa u_k}{\pa a^1} D_{1k} =
\frac{D_{1r}}{g_r} \sum_{k=1}^r g_k\, \frac{\pa u_k}{\pa a^1} =
D_{1r}\, a_1 \frac{W_1}{g_r}.
\eeq
Combining (\ref{J}) and (\ref{Dexpr}) we obtain
\beq
M_\Phi^2 = \left( W_1| \right)^2=
\left( \frac{c_1}{c_2} \right)^2 g_r^2 \prod_{\B \in \Delta_d,\, \B>0} 
a \cdot \B .
\label{MPhi}
\eeq

Upon integrating out the massive adjoint we
relate the scale $\La_H$ with the scale $\La_{YM}$ of the
low-energy $N=1$ $SU(2)$ Yang-Mills theory by
\beq
\La_H^{2 \cdot 2 } = \La_{YM}^{ 3 \cdot 2}/M_\Phi^2 .
\eeq
We finally find from this and (\ref{summaa}), (\ref{MPhi}) that 
the scale matching relation becomes
\beq
\La_{YM}^{ 3 \cdot 2} =g_r^2 \La^{2 h},
\label{reg}
\eeq
where the top Casimir $u_r$ has been rescaled so that we can set $c_1/c_2=1$.

Following the previous discussions and the perturbative 
nonrenormalization theorem for the superpotential, 
we derive the low-energy effective superpotential
\beq
W_L= W_{cl}(g) \pm 2 {\La_{YM}}^3 = W_{cl}(g) \pm 2 g_r \La^{h},
\label{exactW}
\eeq
where the term $\pm 2 {\La_L}^3$ has appeared as a result of the gaugino 
condensation in low-energy $SU(2)$ theory
and $W_{cl}(g)$ is the tree-level superpotential evaluated at the
classical values $a^i(g)$. We will assume that (\ref{exactW}) is the exact 
effective superpotential valid for all values of parameters.

\subsection{Determination of singularities and $N=2$ curves}

The vacuum expectation values of gauge invariants are obtained from $W_L$ 
\beq
\langle u_k \rangle = {\pa W_L \over \pa g_k}=u_k^{cl}(g) 
\pm 2 \La^h \delta_{k, r}.
\label{vevofun}
\eeq
We now wish to show that the expectation values (\ref{vevofun}) parametrize
the singularities of algebraic curves. For this let us introduce
\beq
P_{\cal R}(x,u_k^{cl})={\rm det} (x-\Phi_{\cal R})
\label{charapol}
\eeq
which is the characteristic polynomial in $x$ of order ${\rm dim}\, {\cal R}$ 
where ${\cal R}$ is an irreducible representation of $G$. 
Here $\Phi_{\cal R}$ is a representation 
matrix of ${\cal R}$ and $u_k^{cl}$ are Casimirs built out of $\Phi_{\cal R}$.
The eigenvalues of $\Phi_{\cal R}$ are given in terms of the weights 
$\lambda_i$ of the representation ${\cal R}$. Diagonalizing $\Phi_{\cal R}$
we may express (\ref{charapol}) as
\beq
P_{\cal R}(x,a)=\prod_{i=1}^{{\rm dim}\, {\cal R}} (x-a \cdot \lambda_i),
\eeq
where $a$ is a Higgs v.e.v. vector, the discriminant of which takes the form
\beq
\Delta_{\cal R}
=\left( \prod_{i \neq j} a \cdot (\lambda_i-\lambda_j) \right)^2.
\eeq
It is seen that, for $a$ which is a solution to (\ref{eq1}), 
we have $\Delta_{\cal R}=0$, that is
\beq
P_{\cal R}(x, u_k^{cl}(a))=\partial_x P_{\cal R}(x, u_k^{cl}(a))=0
\label{clasin}
\eeq
for any representation. The solutions of the classical equation of
motion thus give rise to the singularities of the level manifold
$P_{\cal R}(x, u_k^{cl})=0$.

In order to include the quantum effect what we should do is to modify the
top Casimir $u_r$ term so that the gluino condensation in (\ref{vevofun}) is
properly taken into account. We are then led to take a curve
\beq
\tilde P_{\cal R}(x,z,u_k) \equiv
P_{\cal R} \left( x,u_k+\delta_{k,r} \left( z+\frac{\mu}{z}\right) \right) =0,
\label{curve}
\eeq
where $\mu=\La^{2 h}$ and an additional complex variable $z$ has been 
introduced. Let us check the degeneracy of the curve at the expectation
values (\ref{vevofun}), which means to check if
the following three equations hold
\beqa
\tilde P_{\cal R}(x,z,\bra u_k\ket) & =& 0, \\
\partial_x \tilde P_{\cal R}(x,z,\bra u_k\ket) & =& 0, \\
\partial_z \tilde P_{\cal R}(x,z,\bra u_k\ket) & =& 
\left(1-\frac{\mu}{z^2} \right) \pa_{u_r}
\tilde P_{\cal R}(x,z,\bra u_k\ket)=0.
\label{difxi}
\eeqa
The last equation (\ref{difxi}) has an obvious solution $z=\mp \sqrt{\mu}$. 
Substituting this into the first two equations we see that the singularity
conditions reduce to the classical ones (\ref{clasin})
\beqa
\tilde P_{\cal R}(x,\mp \sqrt{\mu},\bra u_k\ket) 
& =& P_{\cal R} \left( x,\bra u_k\ket \mp \delta_{k,r} 2 \sqrt{\mu} \right)
=P_{\cal R} ( x,u_k^{cl} )=0,    \\
\pa_x \tilde P(x,\mp \sqrt{\mu},\bra u_k\ket)
 & =& \pa_x P_{\cal R} \left( x,\bra u_k\ket \mp \delta_{k,r} 
2 \sqrt{\mu} \right)=\pa_x P_{\cal R} ( x,u_k^{cl} )=0.
\eeqa
Thus we have shown that (\ref{vevofun}) parametrize the singularities of
the Riemann surface described by (\ref{curve}) irrespective of the
representation ${\cal R}$.

Let us take the $N=2$ limit by letting all $g_i \rightarrow 0$ 
with the ratio $g_i/g_r$ fixed, then 
(\ref{curve}) is the curve describing the Coulomb phase of 
$N=2$ supersymmetric Yang-Mills theory with ADE gauge groups.
Indeed the curve (\ref{curve}) in this particular form of foliation agrees 
with the one obtained systematically in \cite{MaWa} in view of integrable 
systems \cite{Gor},\cite{NT},\cite{IM}. For $E_6$ and $E_7$ see 
\cite{LW},\cite{WY}. 

Finally we remark that there is a possibility of
(\ref{difxi}) having another solutions besides
$z=\mp \sqrt{\mu}$. If we take the fundamental representation such
solutions are absent for $G=A_r$, and for $G=D_r$ there is a solution
with vanishing degree $r$ Casimir (i.e. Pfaffian),
but it is known that this is an apparent singularity \cite{BL}. For
$E_r$ gauge groups there could exist additional solutions.  We expect that 
these singularities are apparent and do not represent physical massless
solitons.

\section{Superconformal field theories}

We will discuss non-trivial fixed points in our $N=1$ theory characterized
by the microscopic superpotential (\ref{tree}). To find critical points we 
rely on the recent construction of new $N=2$ superconformal field theories
realized at particular points in the moduli space of the Coulomb
phase \cite{AD},\cite{APSW},\cite{EHIY},\cite{EH}. 
At these $N=2$ critical points mutually non-local massless dyons
coexist. Thus the critical points lie on the singularities in the moduli
space which are parametrized by the $N=1$ expectation values (\ref{vevofun})
as was shown in the previous section. This enables us to adjust the
microscopic parameters in $N=1$ theory to the values of $N=2$ non-trivial
fixed points. Doing so in $N=2$ $SU(3)$ Yang-Mills theory Argyres and Douglas
found non-trivial $N=1$ fixed points \cite{AD}. 
We now show that this class of $N=1$
fixed points exists in all ADE $N=1$ theories in general. See \cite{TeYa}
for discussions on AD theories.

Let us start with rederiving $N=2$ critical behavior based on the curve 
(\ref{curve}). An advantage of using the curve (\ref{curve}) is that one can
identify higher critical points and determine the critical exponents 
independently of the details of the curve. 

If we set $z= \mp \sqrt{\mu}$ the condition for higher critical points is
\beq
P_{\cal R}(x,u_k^{cl})=\pa_x^n P_{\cal R}(x,u_k^{cl})=0
\eeq
with $n>2$. Hence there exist higher critical points at 
$u_k=u_k^{sing} \pm 2 \La^h \D_{k,r}$ where $u_k^{sing}$ are the classical 
values of $u_k$ for which the gauge group $H$ with rank larger than one is
left unbroken. The highest critical point corresponding to the unbroken $G$
is located at $u_k=\pm 2 \La^h \D_{k,r}$.

Near the highest critical point the curve (\ref{curve}) behaves as
\beq
u_r+z+{\mu \over z}=c\, x^h+\delta u_k \, x^j,
\label{critcurve}
\eeq
where the second term on the rhs with $j=h-(e_k+1)$ represents a small
perturbation around the criticality at $\delta u_k=0$. 
A constant $c$ is irrelevant
and will be set to $c=1$. Let $u_r=\pm 2\La^h, x=\delta u_k^{1/(h-j)}s$ and
$z \pm \La^h=\rho$, then (\ref{critcurve}) becomes
\beq
\rho \simeq \delta u_k^{h \over 2(h-j)} 
(\mp \La^h)^{1\over 2} (s^h+s^j)^{1\over 2}.
\eeq
We now apply the technique of \cite{EHIY} to verify the scaling behavior of
the period integral of the Seiberg-Witten differential $\lambda_{SW}$.
For the curve (\ref{curve}) it is known that $\lambda_{SW}=xdz/z$. Near the
critical value $z=\mp \sqrt{\mu}$ we evaluate 
\beqa
\oint \lambda_{SW}&=&\oint x{dz \over z} \simeq \oint x d\rho  \CR
&\simeq & \delta u_k^{h+2 \over 2(h-j)} \oint ds 
{hs^h+js^j \over (s^h+s^j)^{1/2}}.
\label{period}
\eeqa
Since the period has the mass dimension one we read off critical exponents
\beq
{2\, (e_k+1) \over h+2}, \hskip10mm k=1,2,\cdots ,r
\eeq
in agreement with the results obtained earlier for $N=2$ ADE Yang-Mills 
theories \cite{EHIY},\cite{EH}.

When our $N=1$ theory is viewed as $N=2$ theory perturbed by the tree-level
superpotential (\ref{tree}) we understand that the mass gap in $N=1$ theory
arises from the dyon condensation \cite{SeWi}. Let us show that the dyon
condensate vanishes as we approach the $N=2$ highest critical point under
$N=1$ perturbation. The $SU(2)\times U(1)^{r-1}$ vacuum in $N=1$ theory 
corresponds to the $N=2$ vacuum where a single monopole or dyon becomes
massless. The low-energy effective superpotential takes the form
\beq
W_m=\sqrt{2} AM\tilde M+\sum_{k=1}^rg_kU_k,
\eeq
where $A$ is the $N=1$ chiral superfield in the $N=2$ $U(1)$ vector multiplet,
$M, \tilde M$ are the $N=1$ chiral superfields of an $N=2$ dyon hypermultiplet
and $U_k$ represent the superfields corresponding to Casimirs $u_k(\Phi)$.
We will use lower-case letters to denote the lowest components of the
corresponding upper-case superfields. Note that $\bra a\ket =0$ in the vacuum
with a massless soliton.

The equation of motion $dW_m=0$ is given by
\beq
-{g_k\over \sqrt{2}}={\pa A\over \pa U_k}M\tilde M, \hskip10mm 1\leq k\leq r
\label{eqmo}
\eeq
and $AM=A\tilde M=0$, from which we have
\beq
{g_k \over g_r}={\pa a / \pa u_k \over
\pa  a / \pa  u_r  }, \hskip10mm 1\leq k\leq r-1,
\eeq
when $\bra a\ket =0$. The vicinity of $N=2$ highest criticality may be
parametrized by
\beq
\bra u_k\ket =\pm 2\La^h \delta_{k,r}+c_k \, \epsilon^{e_k+1}, \hskip10mm 
c_k={\rm constant},
\eeq
where $\epsilon$ is an overall mass scale. From (\ref{period}) one obtains
\beq
{\pa  a \over \pa  u_k } \simeq \epsilon^{{h\over 2}-e_k},
\hskip10mm 1\leq k\leq r,
\eeq
so that
\beq
{g_k \over g_r} \simeq \epsilon^{h-e_k-1} \longrightarrow 0, 
\hskip10mm 1\leq k\leq r-1
\eeq
as $\epsilon \rightarrow 0$. The scaling behavior of dyon condensate
is easily derived from (\ref{eqmo})
\beq
\bra m\ket =\Big( -{g_r \over \sqrt{2} \pa  a / \pa  u_r  }
\Big)^{1/2} \simeq \sqrt{g_r}\, \epsilon^{(h-2)/4} \longrightarrow 0.
\eeq
Therefore the gap in the $N=1$ confining phase vanishes. We thus find that
$N=1$ ADE gauge theory with an adjoint matter with a tree-level superpotential
\beq
W_{\rm crit}= g_{r} u_r(\Phi) 
\label{Wcrit}
\eeq
exhibits non-trivial fixed points. The higher-order polynomial $u_r(\Phi)$
is a dangerously irrelevant operator which is irrelevant at the UV gaussian
fixed point, but affects the long-distance behavior significantly
\cite{KSS}.

\section{Chiral matter multiplets}

In this section we consider $N=1$ gauge theory with $N_f$ flavors of 
chiral matter multiplets $Q^i,\tQ_j$ ($1 \leq i,j \leq N_f$) in addition to
the adjoint matter $\Phi$. Here $Q$ belongs to an irreducible representation 
${\cal R}$ of the gauge group $G$ with the dimension $d_R$ 
and $\tQ$ belongs to the conjugate representation of ${\cal R}$.
We take a tree-level superpotential
\beq
W = \sum_{k=1}^{r} g_{k} u_{k}(\Phi) + 
\sum_{l=0}^{q} {\rm Tr}_{N_f} \, \gamma_l \, \tilde{Q} \Phi_R^l Q,
\label{treem}
\eeq
where $\p_R$ is a $d_R \times d_R$ matrix representation of $\Phi$ in 
${\cal R}$ and $(\gamma_l)_{ij}$, $1 \leq i,j \leq N_f$, are
the coupling constants
and $q$ should be restricted so that $\tilde{Q} \p_R^l Q$ is irreducible
in the sense that it cannot be factored into gauge invariants.
If we set $(\gamma_0)^i_j=m^i_j$ with $[m, m^{\dagger}]=0$, 
$(\gamma_1)^i_j=\sqrt{2} \D^i_j , \, 
(\gamma_l)^i_j=0 $ for $l>1$ and all $g_i=0$, (\ref{treem}) reduces to the
superpotential in $N=2$ supersymmetric Yang-Mills theory with massive
$N_f$ hypermultiplets. 

Let us focus on the classical vacua of the Coulomb phase with $Q=\tilde{Q}=0$ 
and an unbroken $SU(2) \times U(1)^{r-1}$ gauge group symmetry.
The vacuum condition for $\Phi$ is given by (\ref{eqmoratio})
and the classical vacuum takes the form as in the Yang-Mills case
\beq
\p_R = {\rm diag} (a \cdot \lm_1,a \cdot \lm_2, \cdots, a \cdot \lm_{d_R}),
\eeq
where $\lm_i$ are the weights of the representation ${\cal R}$.
In this vacuum, we will evaluate semiclassically 
the low-energy effective superpotential
in the tree-level parameter region where
the Higgs mechanism occurs at very high energies and 
the adjoint matter field $\Phi$ is quite heavy.
Then the massive particles are integrated out and 
we get low-energy $SU(2)$ theory with flavors. 

This integrating-out process
results in the scale matching relation which is essentially the same as the
the Yang-Mills case (\ref{reg}) except that we here have to take into 
account flavor loops. The one instanton factor in high-energy theory 
is given by  $\La^{2 h- l({\cal R}) N_f}$.
Here the index $l({\cal R})$ of the representation ${\cal R}$ is defined by 
$l({\cal R}) \D_{ab}={\rm Tr}( T_a T_b)$ where $T_a$ is
the representation matrix of ${\cal R}$ with root vectors normalized as
$\alpha^2=2$. The index is always an integer \cite{Sla}.
The scale matching relation becomes
\beq
\La_L^{ 3 \cdot 2-l({\cal R}) N_f} =g_r^2 \La^{2 h - l({\cal R}) N_f},
\label{sumatch}
\eeq
where $\La_L$ is the scale of low-energy $SU(2)$ theory with massive flavors.

To consider the superpotential for low-energy $SU(2)$ theory with $N_f$
flavors we decompose the matter representation ${\cal R}$ of $G$ in terms
of the $SU(2)$ subgroup. We have
\beq
{\cal R} =  \bigoplus_{s=1}^{n_{\cal R}} {\cal R}_{SU(2)}^s
\oplus {\rm singlets},
\eeq
where ${\cal R}_{SU(2)}^s$ stands for a non-singlet $SU(2)$ representation.
Accordingly $Q^i$ is decomposed into $SU(2)$ singlets and ${\bf Q}^i_s$
($1 \leq i \leq N_f$, $1 \leq s \leq n_{\cal R}$) 
in an $SU(2)$ representation ${\cal R}_{SU(2)}^s$. $\tilde{Q}_i$ is
decomposed in a similar manner. The singlet components are decoupled in
low-energy $SU(2)$ theory.

The semiclassical superpotential for $SU(2)$ theory with $N_f$ 
flavors is now given by
\beq
W=\sum_{k=1}^{r} g_k u_k^{cl} +
\sum_{s=1}^{n_{\cal R}} 
\sum_{l=0}^{q} (a \cdot \lm_{{\cal R}_s})^l \, 
{\rm Tr}_{N_f} \, \gamma_l \, \tilde{{\bf Q}}_s {\bf Q}_s,
\label{w1}
\eeq
where $\lm_{{\cal R}_s}$ is a weight of ${\cal R}$ which branches to 
the weights in ${\cal R}_{SU(2)}^s$. Here we assume that ${\cal R}$ is
a representation which does not break up into integer spin representations
of $SU(2)$; otherwise we would be in trouble when $\gamma_0=0$.
The fundamental representations of ADE groups
except for $E_8$ are in accord with this assumption.

We now integrate out massive flavors to obtain low-energy $N=1$ $SU(2)$ 
Yang-Mills theory with the dynamical scale $\La_{YM}$. Reading off the flavor
masses from (\ref{w1}) we get the scale matching
\beqa
\La_{YM}^{3\cdot 2} & =&  g_r^2 A(a), \CR
A(a) & \equiv & 
\La^{2 h - l({\cal R}) N_f}
\prod_{s=1}^{n_{\cal R}} \left\{ {\rm det} \left( 
\sum_{l=0}^{q} \gamma_l (a \cdot \lm_{{\cal R}_s})^l
\right)^{l({{\cal R}_{SU(2)}^s})} \right\},
\label{smr}
\eeqa
where $l({{\cal R}_{SU(2)}^s})$ is the index of ${\cal R}_{SU(2)}^s$ which is
related to $l({\cal R})$ through
\beq
l({\cal R}) = \sum_{s=1}^{n_{\cal R}} l({{\cal R}_{SU(2)}^s}).
\eeq
The index of the spin $m/2$ representation of $SU(2)$ is given by
 $m(m+1)(m+2)/6$. 

Including the effect of $SU(2)$ gaugino condensation we finally arrive at 
the effective superpotential for low-energy $SU(2)$ theory
\beq
W_L  = W_{cl}(g) \pm 2 \La_{YM}^3
     = W_{cl}(g) \pm 2g_r \sqrt{A(a)},
\label{Wflavor}
\eeq
The expectation values $\langle u_k \rangle =\pa W_L/ \pa g_k$ are found to be
\beqa
\langle u_j \rangle & = &  u_j^{cl} 
\pm 2 \frac{ \pa \sqrt{A}}{\pa g_j'} , \hskip10mm  1 \leq j \leq r-1, \CR
\langle u_r \rangle & = & u_r^{cl} \pm 2 \left(
\sqrt{A} +g_r \sum_{k=1}^{r-1} \frac{\pa g_k'}{\pa g_r} 
\frac{ \pa \sqrt{A}}{\pa g_k'} \right) \CR
 & =& u_r^{cl} \pm 2 \left( \sqrt{A} -
\sum_{k=1}^{r-1} g_k' \frac{ \pa \sqrt{A}}{\pa g_k'} \right),
\label{vevmat}
\eeqa
where we have set $g_k' = g_k/ g_r$ and used the fact that
$u_k^{cl}$ and $A$ are functions of $g_k'$ since $a_i$ in (\ref{Wflavor}) 
are solutions of (\ref{eq1}) (see also (\ref{eqmoratio})). 

Let us show that the vacuum expectation values (\ref{vevmat}) obey 
the singularity condition for the family of $(r-1)$-dimensional complex 
manifolds defined by ${\cal W}=0$ with coordinates $z, x_1,\cdots ,x_{r-1}$ 
where
\beq
{\cal W} = z+ \frac{A(x_n)}{z}-
\sum_{i=1}^{r} x_i \left( u_i-u_i^{cl}(x_n) \right).
\label{mfd}
\eeq
Here we have introduced the variables $x_i$ ($1\leq i \leq r-1$) instead of
$g_i'$ to express $A(g_n')$ and $u_i^{cl} (g_n')$, $x_r=1$ and $u_i$ are
moduli parameters.
The manifold ${\cal W}=0$ is singular when
\beq
{\pa {\cal W}\over \pa z}=0, \hskip10mm {\pa {\cal W}\over \pa x_i}=0.
\label{Wsing}
\eeq
Then, if we set $ z =\pm \sqrt{A(x_k)}$, $x_k=g_k'$ and $u_j=\bra u_j\ket$
it is easy to show that the singularity conditions are satisfied
\beqa
\left. {\cal W} \right |  & = & \pm 2 \sqrt{A(g_k')} - 
\sum_{i=1}^{r} g_i' \left( \langle u_i \rangle -u_i^{cl}(g_k') \right)=0, \CR
\left. {\pa {\cal W}\over \pa z} \right |  &=& 0,    \CR
\left. \frac{\pa {\cal W}}{\pa x_j} \right | & =&
\pm \frac{1}{\sqrt{A(g_k')}} \frac{\pa A(g_k')}{\pa g_j'}-
\langle u_j \rangle +
\frac{\pa}{\pa g_j'}  \left( \sum_{i=1}^{r} g_i' u_i^{cl} (g_k') \right) \CR
& =& -u_j^{cl} (g_k') + g_r \frac{\pa}{\pa g_j} 
\left( \frac{W_{cl}(g)}{g_r} \right) =0, \hskip10mm 1\leq j\leq r-1.
\eeqa
Thus the singularities of the manifold defined by
${\cal W}=0$ are parametrized by the expectation values $\bra u_k \ket$.

Let us explain how the known curves for $SU(N_c)$ and $SO(2N_c)$
supersymmetric QCD are reproduced from (\ref{mfd}). First we consider 
$SU(N_c)$ theory with $N_f$ fundamental flavors.
Here we denote the degree $i$ Casimir by $u_i$ and correspondingly change
the notations for $x_j$ and $g_j'$. It is shown in \cite{Ki},\cite{KiTeYa}
that 
\beq
A= \La^{2 N_c-N_f}
{\rm det}_{N_f} \left( \sum_{l=0}^q (a^1)^l \gamma_l \right), \hskip10mm
a^1=g_{N_c-1}',
\eeq
and hence (\ref{mfd}) becomes
\beq
{\cal W} = z+ \frac{A(x_{N_c-1})}{z} - 
\sum_{i=2}^{N_c} x_i (u_i-u_i^{cl}(x_n)).
\label{wsu}
\eeq
Since $A$ depends only on $x_{N_c-1}$ we notice that one can eliminate other 
variables $x_1, \cdots ,x_{N_c-2}$ by imposing $\pa {\cal W} / \pa x_j=0$
to get the relation
\beq
u_j^{cl} (x_n) = u_j
\label{impose}
\eeq
for $2 \leq j \leq N_c-2$, and then
\beq
{\cal W} = z+ \frac{A(x_{N_c-1})}{z} -
(u_{N_c} -u_{N_c}^{cl} (x_n)) -x_{N_c-1} (u_{N_c-1}-u_{N_c-1}^{cl}(x_n)).
\label{wsu2}
\eeq

Remember that
\beq
0={\rm det} \left( a^1 - \Phi_{cl} \right) 
=(a^1)^{N_c} -s_2^{cl} (a^1)^{N_c-1}-\cdots -s_{N_c}^{cl},
\label{vacdet}
\eeq
where 
\beq
ks_k+\sum_{i=1}^kis_{k-1}u_i=0, \hskip10mm  u_n={1\over n}{\rm Tr}\, \Phi^n,
\hskip10mm k=1,2,\cdots
\eeq
with $s_0=-1$ and $s_1=u_1=0$. We see with the aid of (\ref{vacdet}) that
\beqa
u_{N_c}^{cl}+ x_{N_c-1} u_{N_c-1}^{cl}  &=&
(u_{N_c}^{cl}-s_{N_c}^{cl} ) +x_{N_c-1} ( u_{N_c-1}^{cl} -s_{N_c-1}^{cl})+
(s_{N_c}^{cl}+ x_{N_c-1} s_{N_c-1}^{cl}) \CR
&=& (u_{N_c}-s_{N_c} ) +x_{N_c-1} ( u_{N_c-1}-s_{N_c-1}) \CR
 && + \left( (x_{N_c-1})^{N_c} -s_2 (x_{N_c-1})^{N_c-1}-
\cdots -s_{N_c-2} \right),
\eeqa
where (\ref{impose}) and the fact that 
$s_{N_c}= u_{N_c} + (\hbox{polynomial of $u_{k}, \;2 \leq k \leq N_c-2$})$
have been utilized. We now rewrite (\ref{wsu2}) as
\beqa
{\cal W} &= & z+ \frac{A(x)}{z} - 
(u_{N_c}+ x u_{N_c-1} ) +(u_{N_c}^{cl}+ x u_{N_c-1}^{cl} ) \CR
 & =& z+ \frac{A(x)}{z} +
x^{N_c} -s_2 x^{N_c-1}-\cdots -s_{N_c},
\label{wsu2f}
\eeqa
where $x_{N_c-1}$ was replaced by $x$ for notational simplicity.
This reproduces the hyperelliptic curve derived in \cite{Ka},\cite{KiTeYa} 
after making a change of variable $y=z-A(x)/z$ and 
agrees with the $N=2$ curve obtained in \cite{HO},\cite{APS},\cite{ArSh} 
in the $N=2$ limit .

Next we consider $SO(2 N_c)$ theory with $2 N_f$ fundamental flavors $Q$.
Following \cite{KiTeYa} we take a tree-level superpotential 
\beq
W=\sum_{n=k}^{N_c-2} g_{2 k} u_{2 k} + g_{2 (N_c-1)} s_{N_c-1}+\lm v
+{1\over 2}\sum_{l=0}^{q} {\rm Tr}_{2 N_f} \, \gamma_l \, Q \p^l Q, 
\label{w4}
\eeq
where
\beqa
u_{2 k} &=& \frac{1}{2 k} {\rm Tr}\, \Phi^{2 k}, 
\hskip10mm 1 \leq k \leq N_c-1,  \CR
v &=& {\rm Pf}\, \Phi=\frac{1}{2^{N_c} N_c !} \E_{i_1 i_2 j_1 j_2 \cdots}
\Phi^{i_1 i_2} \Phi^{j_1 j_2} \cdots  
\eeqa
and
\beq
ks_k+\sum_{i=1}^k i s_{k-i} u_{2i}=0, \hskip10mm s_0=-1, 
\hskip10mm k=1,2,\cdots .
\eeq
According to \cite{TeYa} we have
\beq
(a^1)^2=g_{2(N_c-2)}', \hskip10mm \lm'=2 \prod_{j=2}^{N_c-1} (-i a^j),
\hskip10mm v^{cl}=-g_{2(N_c-2)}' \lm' /2
\label{vcla}
\eeq
and \cite{KiTeYa}
\beq
A= \La^{4(N_c-1)-2 N_f} 
{\rm det}_{2 N_f} \left( \sum_{l=0}^q (a^1)^l \gamma_l \right),
\label{Aa1}
\eeq
and thus
\beq
{\cal W} = z+ \frac{A(x_{N_c-2})}{z} - 
\sum_{i=1}^{N_c-1} x_i (u_{2 i}-u_{2 i}^{cl}(x_n))-x (v-v^{cl}(x_n)),
\label{wso}
\eeq
where $\lm'=\lm/g_{2(N_c-1)}$ was replaced by $x$ and $g_{2i}/g_{2(N_c-1)}$
by $x_i$.

In view of (\ref{Aa1}) we again notice that there are redundant variables 
which can be eliminated
by imposing the condition $\pa {\cal W} / \pa x_j=0$ so as to obtain
\beq
u_{2 j}^{cl} (x_n) = u_{2 j}
\eeq
for $1 \leq j \leq N_c-3$. We then find
\beqa
{\cal W} = z+ \frac{A(x_{N_c-2})}{z} &- &
(u_{2 (N_c-1)} -u_{2(N_c-1)}^{cl} (x_n)) 
-x_{N_c-2} (u_{2(N_c-2)}-u_{2(N_c-2)}^{cl}(x_n))  \CR
&-& x (v-v^{cl}(x_n)).
\label{wso2}
\eeqa
Using $\det (a^1-\Phi_{cl})=0$ we proceed further as in the $SU(N_c)$ case.
The final result reads
\beqa
{\cal W} &= & z+ \frac{A(y)}{z} +
\frac{1}{y} \left( 
y^{N_c} -s_1 y^{N_c-1}-\cdots -s_{N_c-1}y+{ v^{cl}(x_n)}^2 \right) \CR
&& -x (v-v^{cl}(x_n)) \CR
&=& z+ \frac{A(y)}{z} -
\frac{1}{4} x^2 y+y^{N_c-1} -s_1 y^{N_c-2}-\cdots -s_{N_c-1}-v x,
\label{dnfinal}
\eeqa
where we have set $y=x_{N_c-2}$ and used (\ref{vcla}). 
It is now easy to check
that imposing $\pa {\cal W}/\pa x=0$ to eliminate $x$ yields the known curve
in \cite{KiTeYa} which has the correct $N=2$ limit \cite{Ha},\cite{ArSh}.

It should be noted here that adding gaussian variables in (\ref{wsu2f})
and (\ref{dnfinal}) we have
\beqa
{\cal W}_{A_{n-1}} &= & z+ \frac{A(y_1)}{z}+y_1^n-s_2y_1^{n-1}-\cdots -s_n
+y_2^2+y_3^2,   \CR
{\cal W}_{D_{n}} &=& z+ \frac{A(y_1)}{z} -
\frac{1}{4} y_2^2 y_1+y_1^{n-1} -s_1 y_1^{n-2}-\cdots -s_{n-1}-vy_2+y_3^2.
\eeqa
These are equations describing ALE spaces of AD type fibered over ${\bf CP}^1$.
Inclusion of matter hypermultiplets makes fibrations more complicated than
those for pure Yang-Mills theory.
For $A_n$ the result is rather obvious, but for $D_n$ it may be interesting
to follow how two variables $y_1,\, y_2$ come out naturally from (\ref{mfd}).
These variables are traced back to coupling constants $g_{2(n-2)}/g_{2(n-1)},\,
\lambda /g_{2(n-1)}$, respectively, and their degrees indeed agree
$[y_1]=[g_{2(n-2)}/g_{2(n-1)}]=2,\, [y_2]=[\lambda /g_{2(n-1)}]=n-2$.

This observation suggests a possibility that even in the $E_n$ case we may
eliminate redundant variables and derive the desired ALE form of SW geometry
directly from (\ref{mfd}) although the problem certainly becomes non-linear.
The issue is under current investigation.

\section{Conclusions}

We have obtained a low-energy effective superpotential for a phase with a 
single confined photon in $N=1$ gauge theory with an adjoint matter with ADE
gauge groups. The expectation values of gauge invariants built out of
the adjoint field parametrize the singularities of moduli space of the $N=2$
Coulomb phase. The result can be used to derive the $N=2$ curve in the form
of a foliation over ${\bf CP}^1$. According to our derivation 
it is clearly observed
that the quantum effect in the SW curve has its origin in the $SU(2)$ gluino
condensation in view of $N=1$ gauge theory dynamics.

In the last year it has been clarified
how SW geometry of $N=2$ Yang-Mills theory appears
naturally in the context of type II compactification on Calabi-Yau 
threefolds \cite{KLMVW},\cite{LW},\cite{review}.
For gauge groups of ADE type SW geometry is derived from ALE spaces of type
ADE fibered over ${\bf CP}^1$. Furthermore SW curves obtained 
from ALE fibrations take the form of spectral curves for the 
ADE periodic Toda lattice. These Toda
spectral curves are described as foliations over ${\bf CP}^1$. Thus it seems
that ALE fibrations combined with integrable systems provide us with a 
natural point of view for SW geometry and the SW curve in the form of
a foliation over ${\bf CP}^1$ is recognized as a canonical description. 
Our study of ADE confining
phase superpotentials also supports this point of view. It is highly desirable 
to develop such a scheme explicitly for non-simply-laced gauge groups and 
supersymmetric gauge theories with fundamental matter fields.


\vskip10mm

The work of ST is supported by JSPS Research Fellowship for Young 
Scientists. The work of SKY was supported in part by the
Grant-in-Aid for Scientific Research on Priority Area 213 
``Infinite Analysis'', the Ministry of Education, Science and Culture, Japan.
SKY would like to thank the ITP in Santa Barbara for hospitality where a part
of this work was done.
This work was supported in part by the National Science
Foundation under Grant No. PHY94-07194.

\newpage



\begin{thebibliography}{99}

\bibitem{review} For recent reviews, see,
W. Lerche, {\it Introduction to Seiberg-Witten Geometry
and its Stringy Origin}, hep-th/9611190;
C. Vafa, {\it Lectures on Strings and Dualities}, hep-th/9702201;
A. Klemm, {\it On the Geometry behind N=2 Supersymmetric Effective Actions
in Four Dimensions}, hep-th/9705131

\bibitem{SeWi} N. Seiberg and E. Witten, 
Nucl. Phys. {\bf B426} (1994) 19; Nucl. Phys. {\bf B431} (1994) 484

\bibitem{InSe} K. Intriligator and N. Seiberg, 
Nucl. Phys. {\bf B431} (1994) 551

\bibitem{ElFoGiRa} S. Elitzur, A. Forge, A. Giveon and E. Rabinovici,
Phys. Lett. {\bf B353} (1995) 79

\bibitem{ElFoGiRa2} S. Elitzur, A. Forge, A. Giveon and E. Rabinovici,
Nucl. Phys. {\bf B459} (1996) 160

\bibitem{In} K. Intriligator, Phys. Lett. {\bf B336} (1994) 409

\bibitem{ElFoGiInRa} S. Elitzur, A. Forge, A. Giveon, K. Intriligator and 
E. Rabinovici, Phys. Lett. {\bf B379} (1996) 121

\bibitem{TeYa} S. Terashima and S.-K. Yang, 
Phys. Lett. {\bf B391} (1997) 107

\bibitem{Ki}  T. Kitao, {\it SUSY N=2 hyperelliptic curve from N=1 effective
potential}, hep-th/9611097

\bibitem{KiTeYa} T. Kitao, S. Terashima and S.-K. Yang,
Phys. Lett. {\bf B399} (1997) 75

\bibitem{GPR} A. Giveon, O. Pelc and E. Rabinovici, {\it The Coulomb Phase
in N=1 Gauge Theories With a LG-Type Superpotential}, hep-th/9701045

\bibitem{LaPiGi} K. Landsteiner, J.M. Pierre and S.B. Giddings, 
Phys. Rev. {\bf D55} (1997) 2367

\bibitem{MaWa} E.J. Martinec and N.P. Warner, Nucl. Phys. {\bf B459} (1996) 97

\bibitem{LW} W. Lerche and N.P. Warner, {\it Exceptional SW Geometry from
ALE Fibrations}, hep-th/9608183

\bibitem{WY} N.P. Warner and S.-K. Yang, to appear

\bibitem{Ito} K. Ito, {\it Picard-Fuchs Equations and Prepotential in N=2
Supersymmetric $G_2$ Yang-Mills Theory}, hep-th/9703180, to appear in 
Phys. Lett. {\bf B}

\bibitem{Hu} J.E. Humphreys, {\it Reflection Groups and Coxeter Groups},
Cambridge University Press (1990)

\bibitem{InSe2} K. Intriligator and N. Seiberg, 
Nucl. Phys. {\bf B444} (1995) 125.

\bibitem{Gor} A. Gorskii, I. Krichever, A. Marshakov, A. Mironov and 
A. Morozov, Phys. Lett. {\bf B355} (1995) 466

\bibitem{NT} T. Nakatsu and K. Takasaki, Mod. Phys. Lett. {\bf A11} (1996) 157

\bibitem{IM} H. Itoyama and A. Morozov, Nucl. Phys. {\bf B477} (1996) 855;
Nucl. Phys. {\bf B491} (1997) 529

\bibitem{BL} A. Brandhuber and K. Landsteiner, Phys. Lett. {\bf B358} (1995) 73

\bibitem{AD} P.C. Argyres and M.R. Douglas, Nucl. Phys. {\bf B448} (1995) 93

\bibitem{APSW} P.C. Argyres,  M.R. Plesser, N. Seiberg and E. Witten,
Nucl. Phys. {\bf B461} (1996) 71

\bibitem{EHIY} T. Eguchi, K. Hori, K. Ito and S.-K. Yang,
Nucl. Phys. {\bf B471} (1996) 430

\bibitem{EH} T. Eguchi and K. Hori, {\it N=2 Superconformal Field Theories
in 4 Dimensions and A-D-E Classification}, hep-th/9607125

\bibitem{KSS} D. Kutasov, A. Schwimmer and N. Seiberg, 
Nucl. Phys. {\bf B459} (1996) 455

\bibitem{Sla} R. Slansky, Phys. Reports {\bf 79} (1981) 1

\bibitem{Ka} A. Kapustin, Phys. Lett. {\bf B398} (1997) 104

\bibitem{HO} A. Hanany and Y. Oz, Nucl. Phys. {\bf B452} (1995) 283

\bibitem{APS} P.C. Argyres, M.R. Plesser and A.D. Shapere,
Phys. Rev. Lett. {\bf 75} (1995) 1699

\bibitem{ArSh} P.C. Argyres and A.D. Shapere, 
Nucl. Phys. {\bf B461} (1996) 437

\bibitem{Ha} A. Hanany, Nucl. Phys. {\bf B466} (1996) 85

\bibitem{KLMVW} A. Klemm, W. Lerche, P. Mayr, C. Vafa and N. Warner,
Nucl. Phys. {\bf B477} (1996) 746

\end{thebibliography}
\end{document}